# Like My Aunt Dorothy: Effects of Conversational Styles on Perceptions, Acceptance and Metaphorical Descriptions of Voice Assistants during Later Adulthood

JESSIE CHIN[*] and SMIT DESAI[*], University of Illinois at Urbana-Champaign, USA
SHENY LIN, University of Illinois at Urbana-Champaign, USA
SHANNON MEJIA, University of Illinois at Urbana-Champaign, USA

Little research has investigated the design of conversational styles of voice assistants (VA) for adults in their later adulthood with varying personalities. In this Wizard of Oz experiment, 34 middle-aged (50 to 64 years old) and 24 older adults (65 to 80 years old) participated in a user study at a simulated home, interacting with a VA using either formal or informal language. Older adults with higher agreeableness perceived VA as being more likable than middle-aged adults. Middle-aged adults showed similar technology acceptance toward the informal and formal VA, and older adults preferred using informal VA, especially those with low agreeableness. Further, while both middle-aged and older adults frequently anthropomorphized VAs by using human metaphors for them, older adults compared formal VA with professionals (e.g., librarians, teachers) and informal VA with their close ones (e.g., spouses, relatives). Overall, the conversational style showed differential effects on the perceptions of middle-aged and older adults, suggesting personalized design implications.

**239**

## 1 INTRODUCTION

In Voice assistants (VAs), like Amazon Alexa, Google Assistant, and Apple Siri, are becoming increasingly popular in everyday life. Users mainly use them for asking for information, playing games, listening to music, home automation, and for companionship [5, 61]. Due to

---

[*] The first two authors are co-first authors and contributed equally to this work

Effects of Conversational Styles of Voice Assistants during Later Adulthood

VAs' accessible speech-based nature, they are especially popular among older adults [64, 65]. According to a Pew Research report in 2019 [1], about 10 million older adults had a smart speaker in their homes.

Currently, VAs are designed using 'humanness' as a metaphor [17], due to which users attribute VAs with human qualities like intelligence, trust, and likeability. The way the VAs are perceived are crucial to their acceptance and long-term adoption. Since VAs are social actors [50], they are often ascribed a personality by the users. Developing VAs with a human personality is a contentious topic in HCI, with arguments questioning the ethical validity of essentially tricking the users into building a relationship with the VAs [51]. Further, researchers also argue if 'humanness' is even an appropriate metaphor for human-VUI conversation as the current versions of VAs are not sophisticated enough to implement the complexities of human-human conversations [19]. Consequently, there has been an impetus to developing conversational styles—instead of personalities—along simplistic dimensions like formality. In a study with 18 participants [37], researchers found that formal VAs were perceived more positively and accepted than informal ones. Similarly, the use of formality has also been associated with users' self-disclosure of personal information in interactions with conversational interfaces [18]. However, this work has been mainly limited to younger adults and specific scenarios (e.g., driving).

There are various factors affecting how users perceive and use VAs, including age-related and personality-based factors. Both younger and older adults like VAs which are personable, but younger adults place more value on intelligence [52]. On the other hand, older adults perceive likeability to be a critical attribute impacting acceptance [13]. Another factor influencing perceptions of VAs is the personality of the end-users themselves, with users preferring to interact with VAs that they perceive as being similar to them [8]. For example, depending on the context, extraverted users prefer VAs whom they perceive as being extraverted [68].

We take all these threads of research into consideration and designed "RAVA," *R*esearch *A*ssisting *V*oice *A*ssistant, a Wizard of Oz simulated VA capable of interacting with users in formal and informal conversational styles. In this paper, we present results from an experiment with 34 middle aged (50 to 64 years old, M=57.94) and 24 older adults (65 to 80 years old, M=71.75), interacting with a VA using two conversational styles—formal and informal—in a simulated home environment. Our study answers three research questions:

**RQ1:** What factors influence the perceived intelligence, trust, and likeability of VAs using different conversational styles (formal v. informal) during later adulthood?

**RQ2:** What constructs influence the technology acceptance of VAs using different conversational styles (formal v. informal) during later adulthood?

**RQ3:** How do age and conversational style (formal v. informal) influence the metaphorical descriptions of VAs during later adulthood?

With these questions, we seek to understand how conversational styles affect the perceptions, acceptance, and metaphorical descriptions of VAs for users in later adulthood with varying personalities (agreeableness). This study addresses a major gap in the VA literature by investigating the effect of conversational styles on middle-aged and older adults—a demographic often ignored in design research. Our study has implications for the design of future VAs for middle-aged and older adults and is a part of an ongoing effort to develop senior-specific VA design guidelines.

## 2 RELATED WORK

### 2.1 Perceptions of VAs

Currently, VAs are designed with the metaphor of 'humanness' [17]. They are based on human-human conversational model and invoke anthropomorphizing behavior. Due to its ability to talk in a natural-sounding voice, users of VAs often assign human-like attributes to them, such as intelligence, trust, and likeability.

The perceived intelligence of a VA is desirable and an important attribute affecting technology acceptance and intention to continue use [12]. When the perceived intelligence of VAs is lower than the users' expectations, they either stop using the VAs or limit their use to perform basic tasks like setting alarms or checking the weather [48]. VA literature suggests that users begin interactions with VAs with a much higher perception of their intelligence; however, with time, the perceived intelligence reduces. Luger & Sellen [48], in interviews with 14 participants, found that users began interactions with VAs with a high expectation of their intelligence but were disappointed due to the VAs' limitation in understanding context and frequent functional errors. Similarly, Cowan et al. [17], found that one of the main obstacles to VA adoption is user expectations being higher than their perceived intelligence. In another study, Wang et al. [70] found that the perceived intelligence of the VAs decreases with continued use.

Users' perceptions of the likeability of VAs mainly depends on the perceived sociability (the degree to which the users think the agent is nice) that they project. In a study with 26 younger and 25 older adults, researchers found that perceived sociability was critically important for older and younger adults [13]. For older adults, the perceived likeability of VAs is more important than the perceived intelligence. Similarly, in a field study with 15 older adults, Hu et al. [32] found that older adults preferred interacting with the polite voice-based system. However, perceived trust is important across in a variety of contexts, including healthcare. Based on the perceived trust of the VAs, users are more likely to share sensitive health-related information [14].

### 2.2 Metaphors and VAs.

Metaphors, beyond literary embellishments, are essential in generating knowledge and understanding. In this paper, we adopt a constructivist standpoint, building on the influential work of Cameron and Maslen [11] and Lakoff and Johnson's [45] foundational research. By emphasizing the role of metaphors in relating concepts to familiar ones, we explore their significance in articulating ideas, forming perceptions, and understanding social roles.

Metaphors in HCI have a long history and traditionally served educational purposes in Graphical User Interfaces (GUIs), explaining interface elements to users [7]. For example, the popular 'desktop' metaphor comprises of secondary metaphors including files, menus, icons, etc. Conversely, the phenomenon of commercial VAs is nascent and lacks design guidelines and research on metaphors [16]. Current VUIs leverage the human-human conversation model and rely on the "humanness" metaphor [21].

The majority of VA literature focuses on understanding user perception using human metaphors [20]. However, recent literature challenges this knowledge by positing that the mental models of users are more complex. Initially, due to novelty, users might assign human attributes to VAs, but with time the limitations become clear, and the mental model significantly changes. Pradhan et al. [55] found that users do not have a fixed mental model to describe VAs. Their ontological understanding of VAs is contextually fluid and switches from thinking of VAs as being object-like to human-like. In another study, Jung et al. [38], adopted Lakoff and Turner's 'Great Chain of Being' [46] framework to study the effect of designing conversational agents (CAs) embodied by non-human (e.g., book and dog) and human metaphors on crowdsourcing tasks. The researchers found that human metaphors had no significant differences in terms of the workers' motivation, trust, and engagement compared to non-human metaphors, and concluded that designing CAs using non-human metaphors is aligned to the users' mental model and could be a deliberate design decision. Further, Desai & Twidale [19, 20] found that the metaphors participants used to describe VAs were dependent on the context of use. Based on the task they were performing; users would switch from thinking of VAs as friends or family pets. In the present study, we aim to capture and describe these complexities by looking at the metaphorical descriptions of VAs and discussing factors that might affect them.

### 2.3 VA Personalities & Conversation Styles

In HCI, the effect of ascribing human properties is often studied under the Computers are Social Actors (CASA) paradigm [50]. Although CASA is based on how humans unintentionally personify computers due to mindless social behaviors, HCI researchers have taken this paradigm further and designed technologies based on human classification models, such as the Big Five [25], consisting of openness, conscientiousness, extraversion, agreeableness, and neuroticism (OCEAN). The current versions of VAs are not technologically sophisticated enough to project these dimensions [69]; however, simpler models have been developed to design conversational styles, including Braun et al.'s model [8]. This classification model suggested designing VAs along two dimensions—formality and equivalence—and found that users preferred interacting with VAs that they perceived to be similar to them.

Drawing from this background, Khadpe et al. [39] conducted a study focused on deliberately designing chatbots for human-AI collaboration tasks with a focus on warmth and competence dimensions. The findings revealed consistent positive outcomes associated with incorporating high levels of warmth in AI chatbots across various scenarios, while the impact of competence varied based on users' pre-existing expectations. Similarly, Gilad et al. [24] explored the influence of warmth and competence on users' intention to adopt AI systems, finding that warmth played a significantly more influential role than competence in shaping user preferences. However, existing research generally supports the notion of designing AI systems that exhibit both high warmth and high competence [24, 38, 39].

Additionally, a related model developed by Tannen [66] and utilized in CAs literature (e.g., [58, 63]) considers involvement (active participation and overlaps in talking) and considerateness (non-imposing and direct communication) as key factors. However, it is worth noting that this model overlaps with the concept of formality, where high involvement and considerateness are associated with an informal conversation style, while a formal style prioritizes higher clarity and directness [40].

Previous studies investigating the effect of conversational styles—mainly in the context of formality—on user perceptions and behaviors have been mainly limited to chatbots and focused on the preferences of younger adults. A study investigating the effect of the formality of a teacher chatbot on learners found that the learners mimicked the formality style of the teacher chatbot [47]. Another study understanding the effect of formality on users' self-disclosure behavior found a significant effect of formality in users' disclosure of sensitive information, with users trusting formal chatbots more with their personal information [18]. Further, Kim et al. [40] found that the conversational style had a strong effect on how users described the chatbots. Participants in their study described the informal chatbot as "friendly," "kind," and "warm." While the formal chatbot is described as "boring" and "rigorous." However, this study involved a younger user group, and the authors highlighted the need to validate their results with other age groups. In our study, we use these concepts to understand how conversational styles affect the metaphorical descriptions of VAs.

Overall, in the literature, the effect of formality on user perception is well documented. Moreover, practitioners and designers of voice interfaces consider the level of formality to be one of the most crucial design decisions [43]. Consequently, in this paper, we discuss the effect of VA formality on older adults' perceptions. Our study is one of the first attempts to apply these concepts to voice-based interfaces to understand the effect of conversational style on middle-aged and older adults' behavior and perceptions.

## 2.4 Aging and VAs

Aging is signified by several changes, including a decline in physical and cognitive functions, vision and hearing loss, and changes in the environmental or social situation [36]. Due to these changes, older adults often suffer increasing feelings of loneliness augmented by depression and stress [31]. However, previous studies indicate that the successful use of assistive technology can help mitigate these problems and encourage an increase in quality of life, autonomy, and mental health [49].

The natural speech-based nature of VAs makes them suitable for older adults and are marketed to them [64, 65]. Due to VAs' utilizing a conversational system similar to human-human conversations, they provide a less steep learning curve and decrease technology acceptance and adoption barriers, especially compared to screen-based GUIs [60]. VAs provide respite from systemic limitations like sensory and cognitive overload, difficulties with using motor function-reliant input modalities like mouse, touchscreens, and touchpads, lack of perceived self-efficacy arising from perceived complexity of use, and reliance on vision acuity [6, 54]. In a study with 15 older adults, Ziman & Walsh [71] found that older adults found VAs to be easier to learn than screen and keyboard-based GUIs.

VAs are perceived positively by older adults due to their intentional anthropomorphic design [41]. In comparison to younger adults, older adults perceive VAs to have a higher social presence (the degree to which the users could imagine the system to be a real person). In a CASA-based experiment comparing the social presence of VAs for older and younger adults, Chin & Desai [13] found that older adults identified social presence as a critical factor impacting technology adoption, acceptance, and continued use and were more likely to anthropomorphize VAs. In another field study at seven older adults' homes, Pradhan et al. [57] found that VAs helped with feelings of loneliness among older adults and were considered social companions with an emphasis on having someone to interact with.

Similarly, in a sixteen-week deployment of VAs in older adults' homes, Kim & Choudhary [42] found that the participants progressed from only using basic functionalities of the VAs (e.g., setting alarms, asking for weather) to treating them more as digital companions. Additionally, with experience older adults developed resilience to conversational errors and developed strategies (such as note-taking) to navigate functional breakdowns.

Although VAs present several accessibility-based benefits, several limitations of VAs hinder their continued use and adoption. Most importantly, older adults face difficulties remembering and using the invocation commands, dealing with conversational errors, and having reservations about the privacy offered by these devices [42]. Older adults formulate longer queries and often take more time to respond to the VAs compared to younger adults. This results in them facing a much higher rate of conversational errors [42], causing anxiety and impacting technology acceptance. This problem is worsened as older adults are regularly not an active part of design processes, making their preferences unexplored [56]. Due to the accessible nature of VAs, there is an urgent need to rectify these problems and develop VAs rooted in older adults' unique preferences to support older adults' need for healthy aging.

## 3 DESIGNING AND IMPLEMENTING RAVA

### 3.1 Designing formality

The formality of a conversation is often understood intuitively by the listener based on the speaker's manner of expression. However, this intuition-based understanding of casual formality is so general that it is meaningless as an analytic concept [34]. Instead, in linguistics, the emphasis is on a different form of formality described as 'deep formality,' characterized by didactic attention to circumvent ambiguity, resulting from 'context dependence and fuzziness of expressions' [29].

Compared to informal speech, formal speech is more direct and involved. In spoken language, this is exemplified by using the first-person narrative for informal speech and the third-person narrative for formal speech. To quantify these understandings more comprehensively, Heylighen [29] developed a simple metric for measuring formality relying on decreased use of pronouns, verbs, adverbs, and interjections and increased use of nouns, adjectives, prepositions, and articles.

Although popular, this approach measures formality from a very narrow individual word-based method and fails to capture the complexities of discourse by neglecting the overall syntax and the goals of the communication. To counter this, Graesser et al. [27] developed a new metric of formality relying on Coh-metrix measures such as word concreteness (lack of abstract words), cohesion (connection of ideas over multiple sentences), syntactic complexity (signified by the simplicity of words and sentence structure), and narrativity (story-based approach closer to communication with friends). This metric asserted that informal language comprises simpler words and syntax, and higher narrativity.

We used Coh-metrix's definition of formality to develop formal and informal conversational styles. The content of the scripts across the two conversational styles remained the same. Further, both conversational styles were delivered with the same voice and speech engine. As previously discussed, the distinguishing characteristic of a formal conversational style includes high cohesion and low narrativity, word concreteness, and syntactic simplicity. Both scripts were designed with an emphasis on these dimensions. It is also important to note that both versions of RAVA were designed to be polite and competent as previous research has established that users—including older adults in particular—value competence and politeness highly [13, 32].

Since RAVA was designed to help with research tasks, the script comprised four main components: (1) introduction, (2) description of the tasks the participants were to perform with the assistance of RAVA, (3) follow-up questions about the task experience, and (4) debriefing. The formal version of RAVA conversed in the third-person narrative, while informal RAVA used a first-person narrative and the pronoun "I." The formal version of RAVA utilized a simpler syntactic structure typified by shorter cogent sentences and fewer abstract words. Table 1 provides examples from the script of how formal and informal versions of RAVA presented the same information. The full versions of both scripts are included as supplementary materials.

Table 1: Implementation of formal and informal versions of RAVA using Coh-metrix's definition of formality

| Scenario | Formal RAVA | Informal RAVA |
| --- | --- | --- |
| Introduction | Hello, this is RAVA, the voice assistant. RAVA will assist you throughout the study and answer any questions that you may have. What is your name? [After participant responds] Okay, RAVA is pleased to make your acquaintance. | Hey there, my name is RAVA. I am the voice assistant. I am here to help you throughout the study and will answer all your questions. So what's your name? [After participant responds] Nice, I am excited to meet you. |
| Task description | Welcome to the study room. Here, you will play the two games that you practiced. You shall play each of the two games a total of four times. RAVA will instruct you on the order of the games. After RAVA shall ask a few questions after each round, as per practice. To begin, please enter your participant ID and enter task ID as "1." | Welcome to the study room. We will play the two games that we just practiced. You will play each of the two games a total of four times. I will guide you on the order of the games and ask a few questions after each round, like we just practiced. To get started, enter your participant ID and task ID as "1." |
| Post-task question | The 30 second break has concluded. Now, please tell RAVA a story that reminds you of what it was like to play the two games just now. | The 30 second break is over. Now, tell me a story that reminds you of what it was like to play the two games just now. |
| Debrief | Thank you for your participation in this study. You may now exit this room. A researcher will meet you at the door. | Thanks for participating. You can now exit this room and meet the researcher at the door. |

## 3.2 Norming Study on Formality

As a manipulation check, we recruited 145 adults in an online study through Prolific. Following the within-subjects design, participants listened to three audio clips of a formal-tone RAVA and three audio clips of an informal-tone RAVA. Six audio clips and an attention check audio clip were presented in a random order. After listening to each audio clip, participants would rate the formality of RAVA on a scale of 1 (informal) to 7 (formal). Repeated Measures ANOVA was conducted to examine the effect of conversational style (formality) on the perceived ratings. Results showed that participants rated the formal RAVA (M=5.27, SE=0.07) being significantly more formal than the informal RAVA (M=3.99, SE=0.08)(F(1,144)=205.78, p<.001, partial η$^2$ =0.59), confirming the efficacy of our manipulation in formality.

### 3.3  Talking to RAVA

In this study, we used a Wizard of Oz experiment to implement RAVA. In our study setup, the participants interacted with RAVA in the study room of a simulated smart home space (as shown in Fig. 1) without the presence of researchers. By simulating a smart home space and conducting the study without the presence of researchers, the participants could interact with RAVA in a relatively more authentic and realistic manner. This setup allowed for a higher level of immersion and minimized potential biases or influences that could arise from the presence of researchers. Additionally, it provides an opportunity to observe participants' natural behaviors and reactions, enhancing the ecological validity of the study. The smart speaker (Google Nest Mini Gen. 2) was placed on a table near the participants. In order to create the 'illusion' that participants were talking to an AI, we used the Bluetooth functionality of the smart speaker and connected it to Amazon's Text to Speech (TTS) service, Amazon Polly. We chose the female voice 'Joanna' and set the speech engine to neural to project a more human-sounding voice. The smart speaker was operated by a researcher using a Bluetooth computer in another room using a structured script to respond to the participants' commands and responses with either of the two conversational styles—formal or informal—based on the study condition. Participants were informed of the Wizard of Oz nature of the experiment after all data collection was completed.

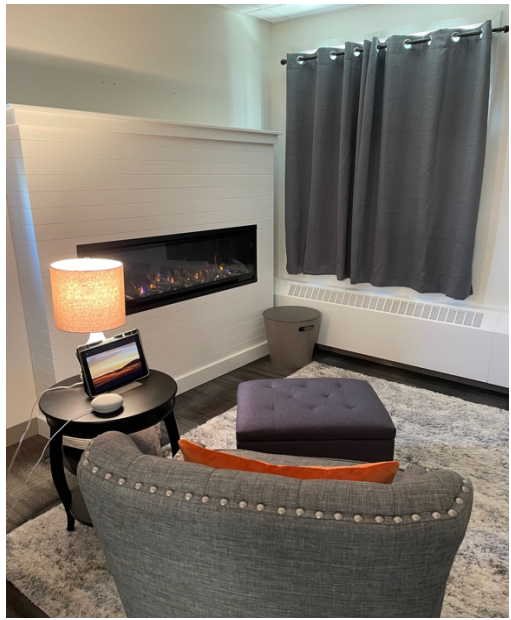

Effects of Conversational Styles of Voice Assistants during Later Adulthood

Fig. 1: A depiction of the study room equipped with a smart speaker, a tablet (to play games), and comfort amenities.

## 4 PRESENT STUDY

### 4.1 Participants

The determination of the sample size was conducted using power analysis to detect a between-subjects difference, assuming a medium to large effect size (d=0.65) [22]. 34 middle-aged adults (M=57.94, 18 Females) and 24 older adults (M=71.75, 13 Females) are recruited in the study. To ensure diversity in our sample, we disseminated study ads through the local social media groups and newsletters, the farmer's market, and local seniors' organizations to recruit participants. Following established guidelines [10, 33], we used age 50 to 64 as middle-aged adults, and age 65 and older ad older adults. One participant from Middle-Aged group self-identified as Black or African American, and the remaining participants were White or Caucasian. Most of the participants completed some years of college degree. Participants were paid $40 for their participation. The demographic information of the participants is summarized in Table 2.

Table 2: Descriptive statistics of the participants

| Variables | Middle-Aged (50-64 years old) | Older (65-80 years old) |
|---|---|---|
| Age (M, SD) | 57.94 (4.22) | 71.75 (4.04) |
| Education | | |
|    High School | 8.8% | 4.2% |
|    1-3 years of college | 11.8% | 16.7% |
|    Associate's degree | 5.9% | 0 |
|    Bachelor degree | 32.4% | 29.2% |
|    Master degree | 35.3% | 41.7% |
|    PhD degree | 5.9% | 8.3% |
| Occupation | | |
|    Management | 5.9% | 4.2% |
|    Business and Financial | 2.9% | 8.3% |
|    Architecture and Engineering | 0 | 4.2% |
|    Community and Social Service | 8.8% | 0 |
|    Education and Library | 26.5% | 25% |
|    Arts, Design, Entertainment, Sports, Media | 2.9% | 8.3% |
|    Healthcare | 5.9% | 4.2% |
|    Protective Service | 2.9% | 4.2% |
|    Cleaning and Maintenance | 2.9% | 0 |
|    Personal Care and Service | 5.9% | 0 |
|    Sales | 5.9% | 0 |
|    Office and Administrative Support | 2.9% | 0 |
|    Production | 2.9% | 4.2% |
|    Military | 0 | 4.2% |
|    Other | 23.5% | 33.3% |

## 4.2 Apparatus & Materials

All participants completed four questionnaires. (1) demographic questionnaire, including questions about age, gender, ethnicity and racial groups, education and occupation;(2) personality measures, using the Ten-Item Personality Inventory (TIPI) [26], which measured the Big Five constructs of personality using ten questions with a 7-point Likert's type scale, including extraversion, agreeableness, conscientious, emotional stability, and openness; (3) perceptions of VA, using the Godspeed questionnaire [3] to measure the perceived likeability, trust, and intelligence of VA. Godspeed questionnaire has been popularly used in studies with voice assistants [62] to measure user perceptions ; (4) technology acceptance, measured by the Unified Theory of Acceptance and Use of Technology [67] (UTAUT; 8 constructs including performance and effort expectancy, attitudes, social influence, facilitating condition, anxiety, self-efficacy, and behavioral intentions to use the technologies, (Cronbach's α = .77 - .94). We included two new contrasts, perceived sociability and social presence as suggested in previous studies in human-robot interaction [28].

## 4.3 Design & Procedure

We followed a 2 (age group: middle age vs. older) x 2 (conversational style: Informal vs. Formal) between-subjects study design. Half of the middle-aged and older adults were assigned to interact with RAVA using a formal conversational style, and the other half of the participants were assigned to interact with RAVA using an informal conversational style in a counterbalanced order.

All participants were invited to a 2-hour study in a simulated smart home space. Following the university-approved COVID-19 human subject study protocol, participants performed three stages of study procedures in different rooms of a home facility. All participants would first complete the demographic and personality questionnaires in the dining room. Participants would then head to the study room to play cognitive games on an iPad for about an hour. While in the study room, the participants were left alone with a Bluetooth smart speaker (Google Nest Mini Gen. 2) equipped with RAVA. During this part of the experiment, the VA would guide participants to play two kinds of cognitive games on the iPad and ask them to share their reflections on their game performance. Following the guidance of RAVA, participants would play the games four times and share their reflections after each game. After the game playing, the human experimenter guided participants to sit in the living room to complete a few questionnaires, including the perceptions and technology acceptance of VA, and go through a semi-structured interview. During the semi-structured interview, participants would share their impressions and thoughts about their experience using the voice assistant and the home environment. Further, the participants were asked to share their metaphorical impressions of RAVA using the prompt, "RAVA is like a…?" Participants were free to provide as many metaphors as they wished. Finally, participants would receive a debriefing (during which the Wizard of Oz nature of the study was revealed) and reimbursement at the end of the 2-hour study session.

## 5 RESULTS

Given the multiple comparisons used in the study, False Discovery Rate (FDR) control method was conducted to reduce Type-1 errors with adjust p-values following Benjamini-Hochberg (BH) procedures (p.adjust function in R). The reported p-values below are adjusted p-values.

### 5.1 RQ1. What factors influence the perceived intelligence, trust, and likeability of CAs using different conversational styles (formal v. informal) during later adulthood?

We conducted linear regressions (lm package in R) to examine the effects of age on perceived intelligence, perceived likeability, and perceived trust of VA with different conversational styles. Following previous literature [13], older adults were found to demonstrate more socially desirable attitudes and prosocial behavior during the interactions with VA than the younger ones. Hence, we chose "agreeableness" in personality measures, as a factor to explain the individual differences in the effects of conversational styles on the perceptions of VAs. The categorical variables of age group and conversational style were examined using contrast testing.

We found a significant interaction effect of age x agreeableness on perceived likeability of VA (see Fig. 2; B=0.67, SE=0.23, t=2.87, p<.05), suggesting that older adults rated VA more likeable than middle-aged adults, especially those with higher agreeableness. There were no other interaction effects (agreeableness x conversational style: B=0.38, SE=0.23, t=1.61, p=0.21; age x conversational style: B=-2.02, SE=2.59, t=-0.78, p=0.44; agreeableness x age x conversational styles: B=0.40, SE=0.47, t=0.85, p=0.44). Interestingly, conversational style did not influence the likeability of VA (B=-2.03, SE=1.30, t=-1.57, p=0.21). For perceived intelligence and trust, we did not find any significant effects of age, formality, agreeableness, and their interactions.

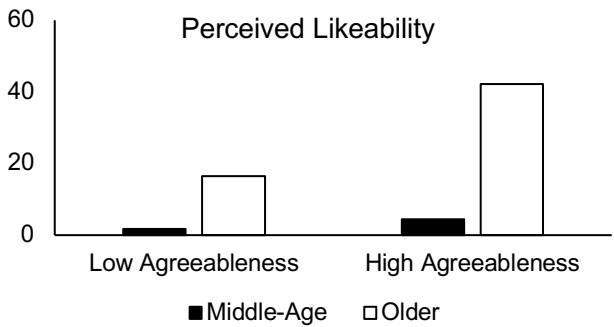

Fig. 2. The marginal differences of the effect of age x agreeableness on perceived likeability

### 5.2 RQ2. What constructs influence the technology acceptance of CAs using different conversational styles (formal v. informal) during later adulthood?

We first conducted correlations to examine the associations among constructs in technology acceptance, including performance expectancy (PE; perceived usefulness), effort expectancy (EE; perceived ease of use), attitudes (ATT), social influence (SI), facilitating condition (FC), self-efficacy (SE), anxiety (ANX), perceived sociability (PS), social presence (PS), behavioral intention (BI), and perceptions toward VA, including perceived likeability (LIK), intelligence (INT) and trust. Results found that adults would more like to continue using VA if performance expectancy, attitudes, social influence, perceived sociability, social presence, intelligence, and likeability of VA are higher (see Table 3).

Table 3: Correlations among the constructs in technology acceptance

| | PE | EE | ATT | SI | FC | SE | ANX | PS | SP | BI | LIK | INT |
|---|---|---|---|---|---|---|---|---|---|---|---|---|
| PE | | | | | | | | | | | | |
| EE | 0.48* | | | | | | | | | | | |
| ATT | 0.81* | 0.54* | | | | | | | | | | |
| SI | 0.48* | 0.19 | 0.41* | | | | | | | | | |
| FC | 0.29* | 0.35* | 0.36* | 0.32* | | | | | | | | |
| SE | -0.10 | 0.20 | -0.10 | -0.17 | -0.04 | | | | | | | |
| ANX | -0.09 | 0.25 | 0.01 | -0.16 | 0.25 | 0.37* | | | | | | |
| PS | 0.48* | 0.34* | 0.56* | 0.27* | 0.10 | -0.02 | -0.18 | | | | | |
| SP | 0.32* | -0.07 | 0.36* | 0.39* | 0.19 | -0.36* | -0.33* | 0.45* | | | | |
| BI | 0.38* | 0.13 | 0.52* | 0.35* | 0.18 | -0.13 | 0.74 | 0.36* | 0.46* | | | |
| LIK | 0.45* | 0.39* | 0.51* | 0.11 | 0.07 | -0.06 | 0.14 | 0.70* | 0.20 | 0.44* | | |
| INT | 0.30* | 0.37* | 0.32* | 0.04 | 0.09 | 0.09 | 0.03 | 0.58* | 0.24 | 0.26* | 0.63* | |
| Trust | 0.10 | 0.32* | 0.16 | -0.25 | 0.05 | 0.04 | 0.08 | 0.41* | 0.01 | 0.13 | 0.61* | 0.49* |

Note: *p<.05

Effects of Conversational Styles of Voice Assistants during Later Adulthood

We then conducted the regression analysis (lm package in R) to examine the effects of age, agreeableness and conversational styles on their behavioral intentions (to continue using VAs). Results found a significant interaction effect of age x conversational style (B=-8.69, SE=3.25, t=-2.67, p<.05) and a signification interaction effect of age x conversational style x agreeableness (B=1.54, SE=0.58, t=2.65, p<.05). As shown in Fig. 3, older adults showed much higher behavioral intentions for informal VA than the formal VA, especially for those with low agreeableness. For middle-aged adults, there were no differences in technology acceptance regardless of their personality traits or conversational styles.

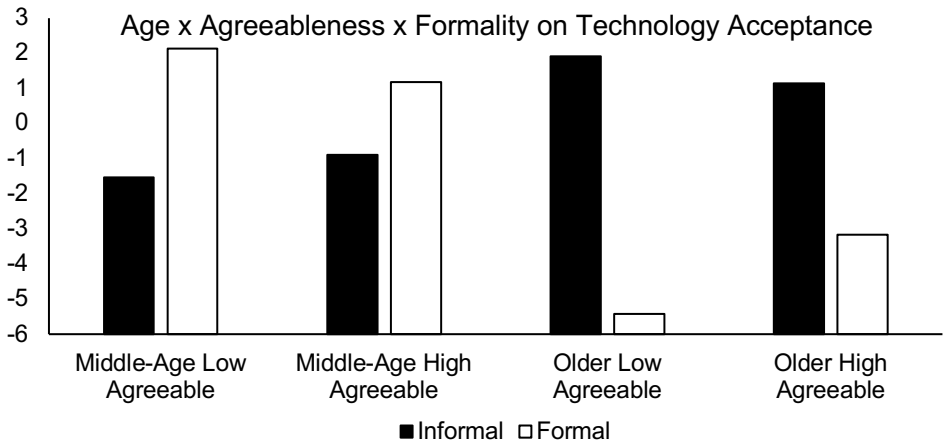

Fig. 3. The effects of age x agreeableness x conversational style on technology acceptance

Overall, we found that middle-aged adults had similar acceptance toward the VAs regardless of their conversational styles. However, older adults would prefer using informal VA, especially if they have lower agreeableness.

## 5.3 RQ3: How do age and conversational style (formal v. informal) influence the metaphorical descriptions of VAs during later adulthood?

To derive our coding scheme, we are inspired by two systems: (1) Lakoff and Turner's 'great chain of being' [46] and (2) Hofstadter's 'cone of consciousness' [30]. In Lakoff and Turner's system, beings or objects are placed on a hierarchical spectrum with inorganic objects at the bottom and God at the top. Similarly, Hofstadter's cone of consciousness places atoms at the bottom and humans (particularly "I") at the top. However, both widely used classification systems lack categorization of technologies in a similar hierarchical structure based on their characteristics or how they are perceived. In our study, we observed participants comparing the VA to not only other humans but also non-human entities like other machines (e.g., Roomba or alarm clocks) or plants (e.g., mushroom). To classify these complex and varied categories of metaphors, we developed a system using definitions of agency, autonomy, and animacy derived from social robotics literature [2, 23, 35].

Two researchers defined the categories and coding schemes of metaphors in terms of agency, autonomy, and animacy (See Table 4). Disagreements were resolved by discussions.

Effects of Conversational Styles of Voice Assistants during Later Adulthood

Agency is operationalized as to what degree the VA is able to act or make decisions on behalf of a user for social purposes. Autonomy is operationalized as to what degree the VA is able to govern and direct one's behavior without external stimulus. Animacy is operationalized as to what degree the VA is alive, living, or able to move around physically. Using agency, autonomy, and animacy, we identified four categories of metaphorical classification of VAs. The first category includes any metaphorical descriptions signifying no agency and no autonomy with varying levels of animacy (in terms of being a living entity). Most of the metaphors mentioned in this category are static living or non-living objects, such as a stone, pencils, trees, etc. The second category includes any metaphorical descriptions signifying no agency, and varying levels of autonomy and partial animacy in terms of being able to perform physical movements. Most of the metaphors mentioned in this category are technologies that can perform movements under the control of others or autonomously to some degree, such as a remote-controlled car, drones, or Roomba. The third category includes any metaphorical descriptions signifying some agency, autonomy, and mixed animacy in terms of physical movement. Most of the metaphors mentioned in this category are actual or fictional intelligent and smart technology, which can conduct some tasks without detailed instructions, such as robot, Alexa, drones, HAL9000 from Space Senior Odyssey. The final category is humanlike expressions having agency, autonomy, and animacy. The metaphors in this category vary from someone participants know to a general human being or a specific group of people, including friends, kids, my old brother, my aunt Dorothy, personal assistant, librarian, etc.

Table 4: Types of metaphors

| Category | Definition | Types | Examples of participant-used metaphors |
|---|---|---|---|
| 1 | No agency, no autonomy, no or partial animacy (in terms of living, not movement) | Static object | Alarm clock, keyboard, music player, mushroom. |
| 2 | No agency, partial animacy in terms of moving around, varying autonomy | Technology that can move | Drones, Roomba |
| 3 | Agency, autonomy, varying animacy | Technology that is intelligent to support cognitive tasks | Robots, Alexa, Siri, HAL9000 from 2001: A Space Odyssey, Rosey the robot in the Jetsons. |
| 4 | Agency, autonomy, animacy | Human or someone | Personal assistant, friend, my aunt Dorothy. |

For the humanlike category, we made further classification based on the intimacy and identifiability of the person they mentioned (Table 5). The first category is about a general human being or a specific type of human being whom the participants are not related to personally and cannot specifically identify. For example, some middle-aged and older adults described VA as a thinker, an annoying 10-year-old, or a very laid-back person. The second category includes comparisons to a more specific someone belonging to a type of group, usually discerned by their occupations, and personal relatability. For example, participants described VA as their assistant, teacher, librarian, or secretary. The third category is about a specific person that participants can personally relate to, such as their close ones or oneself.

Effects of Conversational Styles of Voice Assistants during Later Adulthood

Participants described VA as their friend, their big brother, their spouse, their aunt, and even themselves.

Table 5: Humanlike Metaphors

| Category | Definition | Types | Examples of metaphors |
|---|---|---|---|
| 1 | Someone who is not identifiable and cannot personally relate to. | Any (type of) human | Human, thinker, explorer, a very laid-back person, an annoying 10-year-old. |
| 2 | Someone who belongs to a specific type of persons and may personally relate to. | Professional | Helpers, secretary, task manager, librarian, assistant, instructor, butler, comedian. |
| 3 | Someone who is a specific person in one's personal life or self. | Close ones: Family or friend or self | Friend, my older sister, big brother, my aunt Dorothy, spouse, kids, voice in my head, myself. |

First, we used multilevel modeling to examine the effects of age, conversational styles, and agreeableness on the metaphors generated (percentage of metaphors per type) using lme4 and lmerTest in R [4, 9, 44, 53]. Contrast coding was used to compare the effects among four types of metaphors, including contrast 1 (static object vs. others), contrast 2 (moving technology vs. smart technology and humanlike), and contrast 3 (smart technology vs. humanlike). We found significant effects of the types of metaphors (moving technology vs. smart technology and human: B=0.24, SE=0.06, t=4.35, p<.001; smart technology vs human: B=0.39, SE=0.06, t=7.05, p<.001), suggesting that participants used humanlike metaphors the most, followed by the smart technology, and the other two types, moving technology (conventional movable technology) and static object.

We also found a significant interaction effect of age x conversational style x types of metaphors (B=-0.78, SE=0.22, t=-3.35, p<.01) (see Fig. 7). For middle-aged adults, they used fewer "conventional movable technology" metaphors than smart technology or humanlike metaphors regardless of the informal or formal VA. For older adults, they used much less "conventional movable technology" metaphors than "smart technology and human" metaphors, especially in informal condition. Hence, older adults were much less likely to associate VA with conventional movable machinery in informal conditions. No effects of agreeableness were found.

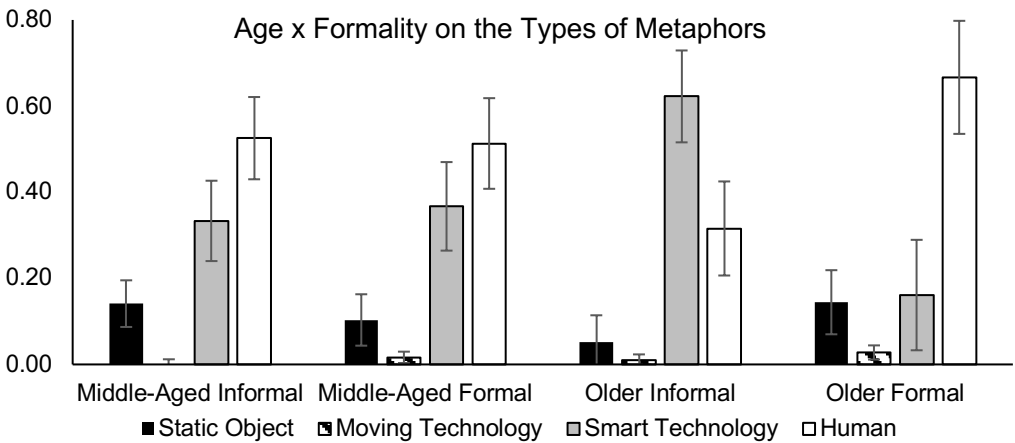

Fig. 4. Effects of age and conversational style on the types of metaphors generated

Since humanlike metaphors were the mostly used metaphors, further analyses were done to examine the use of different humanlike metaphors. We used multilevel modeling to examine the effects of age, conversational styles, and agreeableness on the humanlike metaphors generated (percentage of metaphors per type) using lme4 and lmerTest in R. Contrast coding was also used to compare the effects among three types of metaphors, including contrast 1 (unspecified human vs. others), contrast 2 (professional vs. family and friends). We found a significant interaction effect of age x conversational style x types of humanlike metaphors (B=-1.19, SE=0.32, t=0.37. p<.01). As shown in Figure 5, middle-aged adults did not differ in their association between VA with professionals based on the conversational style. However, older adults generated more professional metaphors when VA was using a formal conversational style than an informal one. A higher frequency of professional metaphors could also explain older adults' association of formal VAs with more humanlike metaphors. On the contrary, for the metaphors relating to close ones, older adults generated much more metaphors relating to close ones (friends and family) when VA was using informal conversational style than the formal one. However, middle-aged adults generated roughly the same number of metaphors for close ones regardless of the conversational styles. No effects of agreeableness were found.

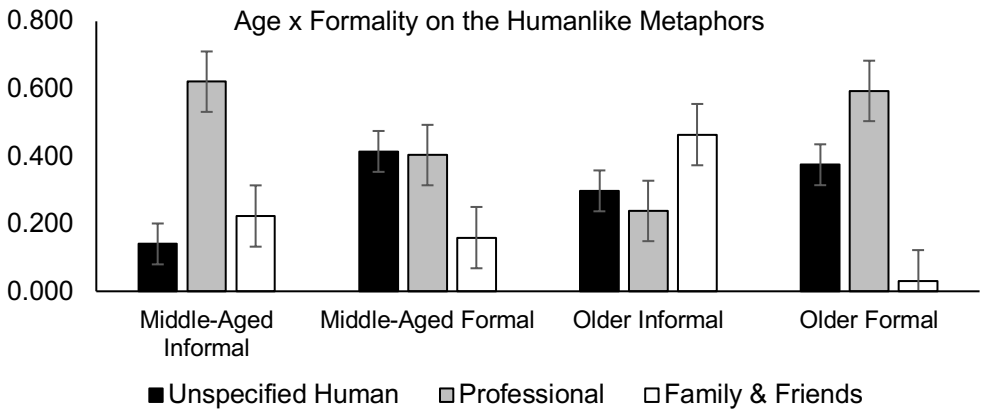

Fig. 5. The effects of age x conversational style on humanlike metaphors

In sum, older adults are more sensitive to the conversational style of VA than the middle-aged adults, especially by relating VA with close ones when using informal conversational style, and professionals when using formal conversational style.

## 6   DISCUSSION

Our study showed that older adults would perceive VA as more likable than middle-aged adults, especially those with higher agreeableness. Regarding technology acceptance, middle-aged adults demonstrated an equivalent willingness to continue using the informal and formal VA. However, older adults showed a higher willingness to use an informal VA, especially those with lower agreeableness. Finally, we found that middle-aged adults tended to use metaphors related to smart technologies or other human beings to describe VAs regardless of the conversational style; so did older adults who interacted with informal VA. However, older adults interacting with formal VA were likelier to use professional human metaphors. Further, older adults described VA as professionals, such as teachers, personal assistants, and librarians, when VA was using a formal conversational style and described VA as close ones, such as their family members or friends or even themselves when VA was using an informal conversational style. Overall, the study concluded how middle-aged and older adults react differently to the informal and formal VA.

**6.1 The social desirability effects**

Agreeableness implies the degree to which a person shows socially desirable behavior, such as being friendly and warm. Our findings suggested that agreeableness is an important personality trait that would determine not only the perceptions (e.g., perceived trust and likability) but also the technology acceptance of VA, especially for older adults. Further, previous studies with VAs indicate that older adults are inclined to seek social companionship from their VAs [42, 56]. As a result, older adults might apply their social desirability toward the VA and try "being nice." For example, a study [13] examined how younger and older adults perceived the sociability (i.e., being pleasant to talk with) of VA when participants were either interviewed by the human experimenter or by the VA itself. The findings using the CASA paradigm suggested that older adults perceived VA as being more pleasant to talk with if they were interviewed by VA than the human experimenter. Younger adults did not show this difference in perceived sociability of VA across the types of interviewers. This finding supplemented by our results suggests that older adults are likelier to try to maintain a friendly relationship with VAs than younger adults. Future studies should investigate whether VA's agreeableness would influence older users' perceptions reciprocally.

**6.2 Like my aunt Dorothy**

In congruence with previous studies investigating the effect of metaphors on user perceptions [20, 38], we found that the conversational style of VAs had a significant effect on the metaphors used by older adults. Older adults were likelier to use closer relationships to describe an informal VA. They had a specific image and used specific names to describe VAs, such as "like my aunt Dorothy," "like my big brother who always listens to me", "seeing a connection to my sisters," "my kid," etc. These metaphors imply that the informal VA was considered to evoke more personal metaphorical associations than the formal VA. As previously discussed, older adults show high proclivity to seek social companionship from their VAs than other age groups. Our results indicate that some degree of companionship could be elicited among older adults by relatively simple formality-based manipulation of conversational style. Since older adults indicated their intention to continue using informal VA than formal VA, a formality-based approach could be an effective tool for Conversational UX designers to positively influence older adults' continued use and perceptions.

In addition to how users attribute human-like metaphors to VAs, another interesting finding is that middle-aged and older adults would use fictional characters to describe VAs. Multiple middle-aged and older adults described VA as HAL9000 from the movie Space

Odyssey, and some described VA as Casper the friendly ghost, Big Brother from 1984, Elmo, or Rosey the Robot. These metaphorical descriptions from early-time film productions could help middle-aged and older adults make sense of VA and develop their norms for interacting with VA. Since human-VA interaction is a relatively novel phenomenon, mainstreamed by the launch of the Amazon Echo Dot in 2014, users do not have a clear mental model of how to have a 'conversation' with their VAs [19]. In this context, fictional metaphors could be used by UX researchers to establish a common ground with the users to discuss VAs. For example, comparisons to HAL9000 or Big Brother could prompt discussions about important user concerns related to VA autonomy and privacy. Similarly, conversational designers could involve end-users to understand how to design VAs that act more like the friendly 'Casper the ghost' and less like the murderous 'HAL9000.'

Another interesting finding is that there were a few participants who saw VA as being like themselves by using the phrase "like myself." One participant even described VA as "the voice in my head." Participants not only metaphorically anthropomorphized the VA but also saw VAs as "extended cognition" [15]. This finding highlights that some users, especially older adults, are heavily influenced by voice-based interfaces, thereby presenting exigent ethical dilemmas, and further challenging the widely accepted 'humanness' metaphor used to design VAs.

**6.3  Design Implications**

Our study finds age-related differences in how middle-aged and older adults perceive VAs. Often adults in their later adulthood are considered to be a monolith, and the heterogeneity of their perceptions is not fully accounted for during the design process [56]. However, our study suggests strong differences in how these adults perceive and interact with VAs. **These age-related differences imply the importance of incorporating a wide variety of personas into the design process.**

Further, when designing VAs for a broader demographic, conversational designers should focus on how to adapt the interaction based on the preferences of the user. Our study shows that adults use fictional metaphors to make sense of their interactions with VUIs. These metaphors could be presented as choices to the end-user, to understand their preferences. Or conversely could be used as system personas to be presented to the end-user. For example, Braun et al. [8] designed in-car voice assistants based on fictional metaphors including, Sherlock Holmes, Sheldon Cooper, Donkey from Shrek, etc. These fictional metaphors were placed along the dimensions of formality and authority and were presented as system personas to the end-user. In this study, we found that the formality dimension, in isolation, evokes unprompted fictional metaphors. Regardless, **fictional metaphors could be leveraged by designers to collect user preferences or used as system personas to avail existing mental models.**

Older adults generally showed higher likeability and trust toward VA than the younger ones (especially for those with high agreeableness). Despite these differential perceptions, both middle-aged and older adults showed equivalent technology acceptance of VA. While middle-aged adults showed a similar willingness to continue using informal and formal VA, the conversational style did matter for older adults, as they preferred informal VA more than formal. Further, older adults were more likely to relate informal VA to their close ones, which partly explains why the informal conversational style made older adults more willing to continue using VA. As a result, we recommend that **when designing VA for older adults, it would be preferable to use informal dialogues to show more intimacy to the older users.**

Additionally, we found agreeableness to be a crucial personality trait determining how middle-aged and older adults perceive voice interfaces. The other design suggestion is to **consider users' personalities in the design of the conversational style of VA.** The personality of the users would influence the perceived intelligence of VA using formal and informal language differently, which perceived intelligence would influence their technology acceptance in general. Some preliminary work [59] suggested the possibility of identifying personality traits through one's texts posted on social media using natural language processing models. Future research could consider expanding this work to recognize the personality traits of the users through their speech-based conversations with VAs and then adjust the conversational style of VA accordingly.

**6.4 Limitations**

There are some limitations of the study. First, although our study aimed to address one of the major gaps in the literature by including both middle-aged and older adults, the recruitment process during the COVID-19 pandemic may have led to a bias toward participants who were able to travel to the university. This could potentially exclude individuals who were more vulnerable to the pandemic or lacked transportation during this time period. The study might unintentionally exclude participants more vulnerable to the pandemic or those lacking transportation during this period. Future studies should consider using home-based studies to recruit participants with varying health conditions across urban and rural areas. Furthermore, it is worth noting that the majority of participants in this study were well-educated and identified as White or Caucasian. Despite the efforts made to recruit a diverse sample, the study sample was not fully representative of the population at large with regard to demographic characteristics. Therefore, it is important for future research to consider the potential impact of such demographic factors of conversational styles on user perceptions. Second, participants were invited to a simulated home environment to interact with VAs for a research study. Although the simulated home environment was designed to replicate in-home studies, the nature of tightly controlled 'research environments' would influence the perceptions, invoking a sense of compliance towards the VA since its primary role was to deliver instructions to the participants and collect reflections on their experiences. Due to this, the participants could have perceived the VA to have a certain sense of authority. Future studies should design VA to be in more fluid roles and see whether adults would generate different role-based metaphorical descriptions or perceptions of the VA. Finally, the VA was delivered using the Google Home Mini smart speaker, which would trigger past personal experiences or background knowledge of this specific device. Indeed, middle-aged adults see VA as smart machinery (e.g., Amazon Alexa, Siri, etc.) as much as a human being, which could be influenced by recognizing the existing properties of the device. Future studies should consider hiding or manipulating the presence of VA.

**7 CONCLUSION**

Effects of Conversational Styles of Voice Assistants during Later Adulthood

Overall, our study suggested that middle-aged and older adults with different personality traits, especially agreeableness, would perceive the voice assistants using formal and informal conversational styles differently. Older adults were sensitive to the conversational styles of voice assistants and regarded them as having different roles accordingly. Interestingly, older adults preferred to use informal conversational style, as it could help them relate to their close ones, such as friends and family members.

Effects of Conversational Styles of Voice Assistants during Later Adulthood